\documentclass[aps, prd, showkeys, showpacs]{revtex4}

\usepackage[small]{caption}
\usepackage{amsmath,amsfonts,amscd,amssymb,epsf, epsfig}\usepackage{graphicx}

\newcommand{\pa}{\partial}

\newcommand{\vep}{\varepsilon}

\begin{document}

\title{  Casimir Effect in Spacetime with Extra Dimensions -- From Kaluza-Klein to Randall-Sundrum Models}

\author{L.P. Teo}\email{ LeePeng.Teo@nottingham.edu.my}\address{Department of Applied Mathematics, Faculty of Engineering, University of Nottingham Malaysia Campus, Jalan Broga, 43500, Semenyih, Selangor Darul Ehsan, Malysia. }



\begin{abstract}
In this article, we   derive the finite temperature Casimir force acting on a pair of   parallel plates   due to a massless scalar field propagating in the bulk of a  higher dimensional brane model. In contrast to previous works which used approximations for the effective masses in deriving the Casimir force, the formulas of the Casimir force we derive are exact formulas.  Our results disprove the speculations that existence of the warped extra dimension can change the sign of the Casimir force, be it at zero or any finite temperature.

\end{abstract}
\maketitle

\section{ Introduction}

Since the advent of string theory, theories of spacetime with extra dimensions become prevalent in physics. The idea of extra dimensional spacetime can be dated back to the work of Kaluza and Klein \cite{1,2}, who tried to propose a theory that can unify classical electrodynamics and gravity. Recently,  intensive investigations on     the Casimir effect in spacetime with extra dimensions are undergoing. In the context of string theory, Casimir effect was studied in \cite{3,4,5,6}. The possible roles played by  Casimir energy as  dark energy or cosmological constant was discussed in \cite{7, 8, 9, 10, 11,12,13,45,56}. The use of Casimir effect in stabilizing extra dimensions were considered in \cite{14,15,16,17,18,19,20,46,54,55}. In the braneworld scenario, Casimir effect was also considered in \cite{47,57,48,49, 52,53,40,50,51}. The influence of the extra dimensions on the Casimir force acting on a pair of parallel plates in macroscopic (3+1)-dimensional spacetime was studied in \cite{13,21,22,23,24,25,26,27,28,29,30,31,32,33,34,58,44}. In the pioneering work of Casimir \cite{42}, it was shown that the Casimir force acting on a pair of   parallel perfectly conducting plates in (3+1)-dimensional spacetime is attractive. It was confirmed later in the work of Mehra \cite{11_02_1} and Brown and Maclay \cite{43} that the thermal correction would not change the sign of the Casimir force. The recent works \cite{13,21,22,23,24,25,26,27,28,29,30,31,32,33,34,58,44} explored the possible influence of the extra dimensions to the magnitude and sign of the Casimir force.   In \cite{13,21,22,23}, the Casimir effect on a pair of parallel plates in spacetime with one extra dimension compactified to a circle was considered. Generalizations to extra dimensional space with more dimensions and more complicated geometries were considered in \cite{24,25,26,27}. Further generalizations to finite temperature Casimir effect were studied in \cite{28,29,30}. In the works \cite{21,22,23,24,25,26,27,28,29,30}, the spacetimes considered are the generalized Kaluza-Klein (KK) models of the form $M^{3+1}\times N^n$ with metric
\begin{equation}\begin{split}\label{eq7_8_3}&ds^2= g^{KK}_{\mu\nu}dx^{\mu}dx^{\nu}=\eta_{\alpha\beta}dx^{\alpha}dx^{\beta} - G_{ab}dx^adx^b,\\
&0\leq \mu,\nu\leq n+3, \;\; 0\leq \alpha,\beta\leq 3, \;\; 4\leq a,b\leq n+3,\end{split}\end{equation} where $\eta_{\alpha\beta}=\text{diag}(1, -1, -1, -1)$ is the usual (3+1)-$D$ metric on the Minkowski spacetime $M^{3+1}$ and   $ds_N^2=G_{ab}dx^adx^b$ is a Riemannian metric on the $n$-dimensional compact internal space $N^n$. In this model, the metric is factorizable. Hence the geometrical structures of the macroscopic manifold $M^{3+1}$ and the internal manifold $N^n$ are independent. It has been concluded that the Casimir force acting on a pair of parallel plates due to a scalar field with homogeneous boundary conditions where Dirichlet  conditions are imposed on both plates ($DD$ conditions)  or Neumann  conditions are imposed on both plates ($NN$ conditions) is always attractive, at either zero or any finite temperature. On the other hand, for mixed boundary conditions where one of the plates assumes Dirichlet boundary condition and the other one assumes Neumann boundary condition ($DN$ conditions), the Casimir force is always repulsive.

In \cite{31,32,33,34}, the spacetime considered is the Randall-Sundrum (RS) brane model. This model was proposed in \cite{36,37} to solve the hierarchy problem between the Planck and the electroweak scale. In this model, the underlying spacetime is a five-dimensional Anti-de Sitter space (AdS$_5$) with background   metric
\begin{equation}\label{eq7_8_1}
ds^2=g^{RS}_{\mu\nu}dx^{\mu}dx^{\nu}=e^{-2\kappa|y|}\eta_{\alpha\beta}dx^{\alpha}dx^{\beta}-dy^2,\hspace{1cm}0\leq \mu,\nu\leq 4,\;\;0\leq \alpha,\beta\leq 3.
\end{equation}This metric is non-factorizable. The extra dimension with coordinate $y$ is  compactified on the orbifold $S^1/\mathbb{Z}_2$. The metric of the underlying Minkowski spacetime depends on the extra dimension through the warp factor $e^{-2\kappa|y|}$, where $\kappa$ determines the degree of curvature of the AdS$_5$ space. There are two types of RS brane models, denoted by RSI and RSII respectively. In RSI, there are two 3-branes with equal and opposite tensions, one invisible and one visible, localized at $y=0$ and $y=\pi R_0$ respectively, where $ R_0$ is the compactification radius of the extra dimension. The $\mathbb{Z}_2$-symmetry is realized by $y\leftrightarrow -y$, $\pi R_0+y\leftrightarrow \pi R_0-y$.  The standard model fields are  localized on the visible brane.   RSII can be considered as a limiting case of RSI where $ R_0\rightarrow \infty$, i.e., one brane is located at infinity. In relation to Casimir effect on parallel plates, RS model was generalized to $(3+n)$-branes with $n$-compact dimensions compactified to an $n$-torus embedded in a $(5+n)$-dimensional spacetime with background metric
\begin{equation}\label{eq7_8_2}
ds^2= e^{-2\kappa|y|}\left(\eta_{\alpha\beta}dx^{\alpha}dx^{\beta}-\sum_{i=1}^n R_i^2d\theta_i^2\right)-dy^2.
\end{equation}In \cite{31} and \cite{33}, it was concluded that the zero temperature Casimir force acting on a pair of parallel plates  in either the (4+1)-D RS model \eqref{eq7_8_1} or its extension \eqref{eq7_8_2} due to a massless scalar field with $DD$ boundary conditions is always attractive. The   methods used in \cite{31} and \cite{33} involve approximations to the tower of masses induced by the   extra dimension $S^1/\mathbb{Z}_2$, and the attractive nature of the Casimir force is not obvious from its analytical expressions. It is also not clear  whether the approximations used in deriving the Casimir force would affect the conclusion about the sign of the Casimir force. Therefore, it is desirable to obtain an exact expression for the Casimir force.

 As mentioned in \cite{11_02_2}, the RS scenario is  the simplest case of  warped geometries. The higher dimensional warped geometries deserve more attention especially in connection with string theory, which asserts that our spacetime should has eleven dimension.   In this article, we consider generalized RS model as in \cite{11_03_1, 11_03_2, 11_03_3, 56,52,53,51} whose background metric is
\begin{equation}\begin{split}\label{eq7_9_9}
ds^2=&g^{RSKK}_{\mu\nu}dx^{\mu}dx^{\nu}=e^{-2\kappa|y|}\left(g^{KK}_{\mu\nu}dx^{\mu}dx^{\nu}\right)-dy^2=e^{-2\kappa|y|}\left(\eta_{\alpha\beta}dx^{\alpha}dx^{\beta}- G_{ab}dx^adx^b\right)-dy^2,\\&0\leq \mu,\nu\leq n+4, \;\; 0\leq \alpha,\beta\leq 3, \;\; 4\leq a,b\leq n+3.\end{split}
\end{equation}Compared to the model \eqref{eq7_8_2}, the internal space now is an arbitrary $n$-dimensional compact manifold with Riemannian metric $ds_N^2=G_{ab}dx^adx^b$. It can be considered as    a KK model \eqref{eq7_8_3} embedded in a RS model \eqref{eq7_8_1}. Therefore we call this model Randall-Sundrum-Kaluza-Klein (RSKK) model.    Our concern here is the Casimir force acting on a pair of parallel plates  rather than the Casimir force acting on the branes which was considered in \cite{56,52,53,51}. As in most of the works about Casimir effect on parallel plates in higher dimensional spacetime, we regard the parallel plates as co-dimension one hyperplanes in the spacetime, and the   field  is assumed to propagate in the bulk. We derive the exact formulas for the finite temperature Casimir force   and show that   warped extra dimensions cannot change the attractive or repulsive nature of the Casimir force. More precisely, it will be concluded that for $DD$ or $NN$ boundary conditions, the Casimir force is always attractive; whereas for $DN$ boundary conditions, the Casimir force is always repulsive.

The units used are such that $\hbar=c=k_B=1$.

\section{Casimir force on parallel plates in Randall-Sundrum-Kaluza-Klein models}
In this section, we derive the Casimir force acting on a pair of parallel plates in the RSKK model with background metric \eqref{eq7_9_9} due to a   scalar field $\Psi(x,y)$  of mass $m$ with equation of motion
\begin{equation}\label{eq7_9_10}
\left(\frac{1}{\sqrt{\left|g^{RSKK}\right|}}\sum_{\mu=0}^{n+4}\sum_{\nu=0}^{n+4}\pa_{\mu}\sqrt{\left|g^{RSKK}\right|}\left(g^{RSKK}\right)^{\mu\nu}\pa_{\nu}+m^2\right)\Psi(x,y)=0.
\end{equation} Using separation of variables,
\begin{equation*}
\Psi(x,y)=\varphi(x)\psi(y),
\end{equation*}the equation of motion \eqref{eq7_9_10} for $y\geq 0$ is equivalent to the following two equations:
\begin{align}\label{eq7_9_11}
&e^{(n+2)\kappa y}\frac{d}{dy}\left(e^{-(n+4)\kappa y}\frac{d\psi(y)}{dy}\right)-m^2e^{-2\kappa y}\psi(y)=-m_{\text{eff}}^2\psi(y),\\\label{eq7_9_11_1}
&\left(\frac{1}{\sqrt{\left|g^{KK}\right|}}\sum_{\mu=0}^{n+3}\sum_{\nu=0}^{n+3}\pa_{\mu}\sqrt{\left|g^{KK}\right|}\left(g^{KK}\right)^{\mu\nu}\pa_{\nu}+m_{\text{eff}}^2\right)\varphi(x)=0.
 \end{align} Eq. \eqref{eq7_9_11_1} is the   equation of motion   for a   scalar field with effective mass $m_{\text{eff}}$ in the KK spacetime. For the first equation in \eqref{eq7_9_11}, the general solutions  can be expressed in terms of the Bessel functions  of the first and second kinds $J_{\nu}(z)$ and $Y_{\nu}(z)$:
\begin{equation}\label{eq7_9_12}
\psi(y)=C_1\psi_{\text{I}}(y)+C_2\psi_{\text{II}}(y)=e^{\frac{(n+4)}{2}\kappa y}\left( C_1 J_{\nu}\left(\frac{m_{\text{eff}}}{\kappa}e^{\kappa y}\right)+C_2Y_{\nu}\left(\frac{m_{\text{eff}}}{\kappa}e^{\kappa y}\right)\right),
\end{equation}where
$$\nu =\sqrt{\left(\frac{n+4}{2}\right)^2+\left(\frac{m}{\kappa}\right)^2}.$$In the massless case, there is an additional solution given by $\psi(y)=$constant and $m_{\text{eff}}=0$.

We first consider the RSKKI model, where two $(3+n)$-branes are localized at $y=0$ and $y=\pi R_0$ respectively. In this case,  boundary conditions   imposed on the branes at $y=0$ and $y=\pi  R_0$ give rise to discrete spectrums of $m_{\text{eff}}^2$. There are various possibilities \cite{40,39,41} of boundary conditions. We choose the one considered in \cite{31,33}, where Neumann boundary conditions are imposed on the branes, i.e., $\left.\pa_y\psi(y)\right|_{y=0}=\left.\pa_y\psi(y)\right|_{y=\pi R_0}=0$. In the massless case, the constant functions satisfy this boundary conditions. The general solution \eqref{eq7_9_12}   satisfies the boundary conditions at $y=0$ and $y=\pi R_0$ if and only if
\begin{equation}\label{eq7_9_16}\begin{split}
C_1\psi_{\text{I}}'(0)+C_2\psi_{\text{II}}'(0)=0,\\
C_1\psi_{\text{I}}'(\pi R_0)+C_2\psi_{\text{II}}'(\pi R_0)=0.
\end{split}
\end{equation}This gives a nontrivial solution to $\psi(y)$ if and only if
\begin{equation}\label{eq7_9_14}
\Delta(m_{\text{eff}})= \psi_{\text{I}}'(0;m_{\text{eff}})\psi_{\text{II}}'(\pi R_0; m_{\text{eff}})-\psi_{\text{II}}'(0;m_{\text{eff}})\psi_{\text{I}}'(\pi R_0; m_{\text{eff}})=0.
\end{equation} For simplicity,   we only consider the massless case, i.e., the $m=0$ case from now on.  In this case, $$\nu=\frac{n+4}{2},$$ and \eqref{eq7_9_14} becomes
\begin{equation}\label{eq7_13_2}
\Delta(m_{\text{eff}})= m_{\text{eff}}e^{\frac{n+6}{2}\pi\kappa R_0}\left\{J_{\frac{n+2}{2}}\left(\frac{m_{\text{eff}}}{\kappa}\right)Y_{\frac{n+2}{2}}\left(\frac{m_{\text{eff}}}{\kappa}e^{\kappa \pi R_0}\right)-
Y_{\frac{n+2}{2}}\left(\frac{m_{\text{eff}}}{\kappa}\right)J_{\frac{n+2}{2}}\left(\frac{m_{\text{eff}}}{\kappa}e^{\kappa \pi R_0}\right)\right\}=0.
\end{equation}Multiplying eq. \eqref{eq7_9_11} with $e^{-(n+2)\kappa y} \psi(y)$ and integrating over $y$ from $y=0$ to $y=\pi R_0$, we find that the effective masses $m_{\text{eff}}$ should satisfy $m_{\text{eff}}^2\geq 0$.  Therefore we only consider the real solutions of \eqref{eq7_13_2}. One can verify that if $m_{\text{eff}}$ is a solution of \eqref{eq7_13_2}, so is $-m_{\text{eff}}$. However, $m_{\text{eff}}$ and $-m_{\text{eff}}$  give rise to linearly dependent solutions of $\psi(y)$. Therefore we only need to consider the positive solutions of \eqref{eq7_13_2}.
Let   $m_{\text{eff},0}=0$ be the effective mass corresponding to the  solution $\psi_0(y)$ = constant and   let $m_{\text{eff},1} < m_{\text{eff},2}<\ldots$ be all the positive solutions of \eqref{eq7_13_2}.    Denote by $\psi_q(y)$ a nontrivial solution to \eqref{eq7_9_11} with $m_{\text{eff}}=m_{\text{eff},q}$ which satisfies the Neumann boundary conditions at $y=0$ and $y=\pi R_0$.

\begin{figure}[h]\centering \epsfxsize=.45 \linewidth
\epsffile{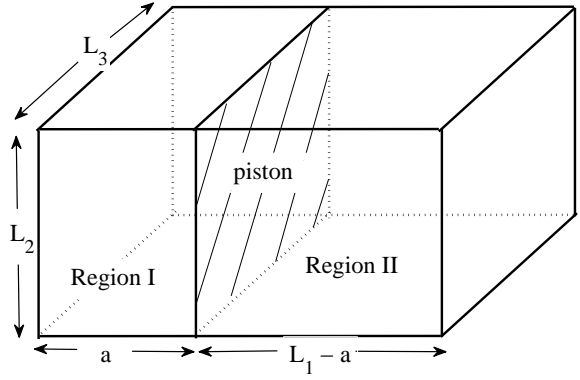}\caption{\label{f1} A
rectangular piston.}\end{figure}

To find the Casimir force acting on  a pair of parallel plates, we use the piston approach \cite{35} (see FIG. \ref{f1}).
 The Casimir force acting on the piston located at $x^1=a$ is given by
\begin{equation}\label{eq7_9_3}
F_{\text{Cas}}^{\text{piston} }(a;L_1;T)= -\frac{\pa}{\pa a}\left(E_{\text{Cas}}^{\text{cavity} }(a;T)+E_{\text{Cas}}^{\text{cavity} }(L_1-a;T)\right),
\end{equation}where $E_{\text{Cas}}^{\text{cavity} }(L;T)$ is the Casimir energy in a cavity of the form $[0,L]\times[0,L_2]\times[0, L_3]\times N^n \times S^1/\mathbb{Z}_2$ defined by
\begin{equation}\label{eq7_9_4}
E_{\text{Cas}}^{\text{cavity}}(L;T)=\frac{1}{2}\sum \omega +T\sum \log\left(1-e^{-\omega/T}\right).
\end{equation}The summation runs through all $\omega$ which are the  eigenfrequencies of the field $\Psi(x,y)$ satisfying the equation of motion  \eqref{eq7_9_10}, with appropriate boundary conditions on the boundary of the cavity. By letting $L_1\rightarrow\infty$, we obtain the Casimir force acting on a pair of parallel plates embedded orthogonally inside an infinitely long rectangular cylinder \cite{38}:
\begin{align}\label{eq9_16_1}
F_{\text{Cas}}^{\parallel }(a;L_1;T)= -\lim_{L_1\rightarrow \infty}\frac{\pa}{\pa a}\left(E_{\text{Cas}}^{\text{cavity} }(a;T)+E_{\text{Cas}}^{\text{cavity} }(L_1-a;T)\right).
\end{align}

 In this article, we impose Dirichlet boundary conditions on the walls $x^2=0, x^2=L_2, x^3=0, x^3=L_3$ of the rectangular cylinder which are transversal to the $x^1$ direction. On the $x^1$ direction, we consider different combinations of boundary conditions. A complete set of independent solutions to the equation \eqref{eq7_9_11_1} with $m_{\text{eff}}=m_{\text{eff},q}$ is given by
\begin{equation}\label{eq7_9_15}
\varphi_{k,\boldsymbol{j},l,q}(x) = e^{-i\omega_{k,\boldsymbol{j},l,q}t}f_k(x^1)\sin\frac{\pi j_2 x^2}{L_2}\sin\frac{\pi j_3 x^3}{L_3}\Phi_l(x^4, \ldots, x^{3+n}), \hspace{0.5cm} k, l\in \mathbb{N}_0, \boldsymbol{j}=(j_2,j_3)\in\mathbb{N}^2,
\end{equation}where for $k=0,1,2,\ldots$,
\begin{equation*}
\begin{split}
f_k(x^1) =& \sin\frac{\pi (k+1)x^1}{L},\hspace{1cm}
f_k(x^1)= \cos\frac{\pi kx^1}{L},\hspace{1cm}
f_k(x^1)=\sin \frac{\pi\left(k+\frac{1}{2}\right)x^1}{L}
\end{split}
\end{equation*}respectively for $DD, NN$ and $DN$ boundary conditions.  For $l=0,1,2,\ldots$, $\Phi_l(x^{4}, \ldots, x^{3+n})$ is an eigenfunction of the Laplace operator $\frac{1}{\sqrt{G}}\pa_a \sqrt{G}G^{ab}\pa_b $ on $N^n$ with eigenvalue $\lambda_{N,l}^2$. By convention, $\lambda_{N,0}^2=0$ corresponds to the constant functions on $N^n$.  The eigenfrequency of $\Psi_{k,\boldsymbol{j},l,q}(x,y)=\varphi_{k,\boldsymbol{j},l,q}(x)\psi_q(y)$ is \begin{equation*}
\omega_{k,\boldsymbol{j},l,q}=\sqrt{\left(\frac{\pi(k+\chi)}{L}\right)^2+\lambda_{\Omega,\boldsymbol{j}}^2+\lambda_{N,l}^2+m_{\text{eff},q}^2},\hspace{1cm}\lambda_{\Omega,\boldsymbol{j}}^2:=\left(\frac{\pi j_2}{L_2}\right)^2+\left(\frac{\pi j_3}{L_3}\right)^2,
\end{equation*}where $\chi=1, 0 , 1/2$ for $DD, NN$ and $DN$ boundary conditions respectively.

  The finite temperature Casimir force \eqref{eq9_16_1} can be calculated in the same way as in \cite{29, 30}, which gives
\begin{equation}\label{eq7_9_17}\begin{split}
&F^{ \parallel}_{\text{Cas}}(a;T) =-(-1)^{2\chi}T\sum_{\boldsymbol{j}\in\mathbb{N}^2}\sum_{l=0}^{\infty}\sum_{p=-\infty}^{\infty}\sum_{q=0}^{\infty}\frac{ \sqrt{\lambda_{\Omega,\boldsymbol{j}}^2+\lambda_{N,l}^2+m_{\text{eff},q}^2+(2\pi pT)^2}}{\exp\left(2a\sqrt{\lambda_{\Omega,\boldsymbol{j}}^2+\lambda_{N,l}^2+m_{\text{eff},q}^2+(2\pi pT)^2}\right)-(-1)^{2\chi}}.\end{split}
\end{equation}The sum of the terms with $q=0$ is the finite temperature Casimir force acting on a pair of parallel plates in the KK model due to a massless scalar field \cite{29,30}.
Notice that each summand in the summation of \eqref{eq7_9_17} is positive. Therefore, the sign of the Casimir force is governed by the factor $-(-1)^{2\chi}$ in front of the summation, which is negative for $\chi=0,1$ and positive for $\chi=1/2$. As a result, we find that the Casimir force is always attractive at any temperature for $DD$  and $NN$ boundary conditions, but always repulsive for $DN$ boundary conditions. This shows that the warped extra dimension cannot change the sign of the Casimir force, but it increases the strength of the Casimir force.

For the influence of the internal extra dimension, notice that the sum of the terms with $l=0$ in \eqref{eq7_9_17} corresponds to the Casimir force in the absence of the internal space $N^n$.  Using again the fact that each summand of the summation in \eqref{eq7_9_17} is positive, one finds  that the extra dimensions enhance the magnitude of the Casimir force. Moreover, the Casimir force   becomes stronger in the presence of more extra dimensions.  When the size of the internal manifold shrinks to zero, only the terms with $l=0$ in \eqref{eq7_9_17} give a nonzero limit:
\begin{equation*}\begin{split}
F^{ \parallel,RSI}_{\text{Cas}}(a;T)=&-(-1)^{2\chi}T\sum_{\boldsymbol{j}\in\mathbb{N}^2} \sum_{p=-\infty}^{\infty}\sum_{q=0}^{\infty}\frac{ \sqrt{\lambda_{\Omega,\boldsymbol{j}}^2 +m_{\text{eff},q}^2+(2\pi pT)^2}}{\exp\left(2a\sqrt{\lambda_{\Omega,\boldsymbol{j}}^2 +m_{\text{eff},q}^2+(2\pi pT)^2}\right)-(-1)^{2\chi}},\end{split}
\end{equation*}which is the finite temperature Casimir force acting on a pair of parallel plates in the RSI model \eqref{eq7_8_1}.

As was proved in \cite{29, 30}, when $L_2=L_3\gg a$, the leading term of the Casimir force \eqref{eq7_9_17} is of order $L_2L_3$. Divide \eqref{eq7_9_17} by $L_2L_3$ and take the limit where $L_2=L_3\rightarrow\infty$, one finds that the Casimir force density acting on a pair of infinite parallel plates in RSKKI model is
\begin{equation}\label{eq7_15_1}
\begin{split}
 \mathcal{F}_{\text{Cas}}^{\parallel }(a;T) =&-\frac{T}{2\pi^{\frac{3}{2}}a^{\frac{3}{2}}}\sum_{k=1}^{\infty}
\sum_{l=0}^{\infty}\sum_{p=-\infty}^{\infty}\sum_{q=0}^{\infty}e^{2\pi i k\chi}\left(\frac{\sqrt{\lambda_{N,l}^2+(2\pi p T)^2+m_{\text{eff},q}^2}}{k}\right)^{\frac{3}{2}}K_{\frac{3}{2}}\left(2ka \sqrt{\lambda_{N,l}^2+(2\pi p T)^2+m_{\text{eff},q}^2}\right)\\
&-\frac{T}{2\pi^{\frac{3}{2}}a^{\frac{1}{2}}}\sum_{k=1}^{\infty}\sum_{l=0}^{\infty}\sum_{p=-\infty}^{\infty}\sum_{q=0}^{\infty} e^{2\pi i k\chi}\frac{\left(\sqrt{\lambda_{N,l}^2+(2\pi p T)^2+m_{\text{eff},q}^2}\right)^{\frac{5}{2}}}{\sqrt{k}}K_{\frac{1}{2}}\left(2ka \sqrt{\lambda_{N,l}^2+(2\pi p T)^2+m_{\text{eff},q}^2}\right),
\end{split}
\end{equation}
Here $K_{\nu}(z)$ is the modified Bessel function of the second kind. The sum of the $l=p=q=0$ term is understood as \begin{equation*}
\lim_{m\rightarrow 0}\left(-\frac{T}{2\pi^{\frac{3}{2}}a^{\frac{3}{2}}}\sum_{k=1}^{\infty}e^{2\pi ik\chi}\left(\frac{m}{k}\right)^{\frac{3}{2}}K_{\frac{3}{2}}\left(2kam\right)-\frac{T}{2\pi^{\frac{3}{2}}a^{\frac{1}{2}}}\sum_{k=1}^{\infty}e^{2\pi ik\chi} \frac{m^{\frac{5}{2}}}{k^{\frac{1}{2}}} K_{\frac{1}{2}}\left(2kam\right)\right)
=\begin{cases}
-\frac{\zeta_R(3)T}{8\pi a^3}, \hspace{0.5cm}&\text{if}\;\; \chi=0,1,\\
\frac{3\zeta_R(3)T}{32\pi a^3}, \hspace{0.5cm}&\text{if}\;\; \chi=1/2.
\end{cases}
\end{equation*}

Taking the zero temperature limit of \eqref{eq7_9_17},
we find that the zero temperature Casimir force acting on a pair of parallel plates in the RSKKI model is
\begin{equation}\label{eq7_16_1} \begin{split}
F^{ \parallel}_{\text{Cas}}(a;0)=&-\frac{1}{2\pi a}\sum_{k=1}^{\infty}\sum_{\boldsymbol{j}\in \mathbb{N}^2}\sum_{l=0}^{\infty}\sum_{q=0}^{\infty}e^{2\pi i k\chi} \frac{ \sqrt{\lambda_{\Omega,\boldsymbol{j}}^2+\lambda_{N,l}^2+m_{\text{eff},q}^2}}{k }K_1\left(2ka \sqrt{\lambda_{\Omega,\boldsymbol{j}}^2+\lambda_{N,l}^2+m_{\text{eff},q}^2}\right)\\&-\frac{1}{\pi}\sum_{k=1}^{\infty}\sum_{\boldsymbol{j}\in \mathbb{N}^2}\sum_{l=0}^{\infty}\sum_{q=0}^{\infty} e^{2\pi i k\chi}\left(\lambda_{\Omega,\boldsymbol{j}}^2+\lambda_{N,l}^2+m_{\text{eff},q}^2\right)K_0\left(2ka \sqrt{\lambda_{\Omega,\boldsymbol{j}}^2+\lambda_{N,l}^2+m_{\text{eff},q}^2}\right).\end{split}
\end{equation}
In the limit where $L_2L_3$ is large, we obtain the zero temperature Casimir force density acting on  a pair of infinite parallel plates:
\begin{equation}\label{eq7_15_4}
\begin{split}
\mathcal{F}_{\text{Cas}}^{\parallel }(a;0)=& -\frac{3}{8\pi^2a^2}\sum_{k=1}^{\infty}\sum_{l=0}^{\infty} \sum_{q=0}^{\infty}e^{2\pi i k\chi} \left(\frac{\sqrt{\lambda_{N,l}^2+m_{\text{eff},q}^2}}{k}\right)^2K_2\left(2ka\sqrt{\lambda_{N,l}^2+m_{\text{eff},q}^2}\right)\\&-\frac{1}{4\pi^2a}\sum_{k=1}^{\infty}\sum_{l=0}^{\infty}
\sum_{q=0}^{\infty} e^{2\pi i k\chi}\frac{\left( \sqrt{\lambda_{N,l}^2+m_{\text{eff},q}^2}\right)^{3}}{k}K_1\left(2ka\sqrt{\lambda_{N,l}^2+m_{\text{eff},q}^2}\right).
\end{split}
\end{equation}The sum of the  terms in \eqref{eq7_15_4} with $l=q=0$ should be understood as
\begin{equation*}
\lim_{m=0}\left(-\frac{3}{8\pi^2a^2}\sum_{k=1}^{\infty} e^{2\pi i k\chi} \frac{m^2}{k^2}K_2(2kam)-\frac{1}{4\pi^2a}\sum_{k=1}^{\infty} e^{2\pi i k\chi} \frac{m^3}{k}K_1\left(2kam\right)\right)=\begin{cases}
-\frac{\pi^2}{480 a^4},  \hspace{0.5cm}&\text{if}\;\; \chi=0,1\\
\frac{7\pi^2}{3840 a^4}, \hspace{0.5cm}&\text{if}\;\; \chi=1/2
\end{cases},
\end{equation*}
which is the the zero temperature Casimir force density $\mathcal{F}_{\text{Cas}}^{\parallel, 3D}(a;0)$ acting on a pair of infinite parallel plates in the $(3+1)$-dimensional Minskowski spacetime due to a massless scalar field. In the limit of vanishing   internal space $N^n$, we obtain the zero temperature Casimir force density acting on a pair of infinite parallel plates in the RSI model:
\begin{equation}\label{eq7_15_2}
\begin{split}
\mathcal{F}_{\text{Cas}}^{\parallel,RSI}(a;0)=&\mathcal{F}_{\text{Cas}}^{\parallel, 3D}(a;0)-\frac{3}{8\pi^2a^2}\sum_{k=1}^{\infty}  \sum_{q=1}^{\infty}e^{2\pi i k\chi} \left(\frac{ m_{\text{eff},q} }{k}\right)^2K_2\left(2ka m_{\text{eff},q} \right)\\&-\frac{1}{4\pi^2a}\sum_{k=1}^{\infty}
\sum_{q=1}^{\infty} e^{2\pi i k\chi}\frac{ m_{\text{eff},q}^{3}}{k}K_1\left(2ka m_{\text{eff},q }\right).
\end{split}
\end{equation} By using the approximation
\begin{equation}\label{eq7_15_3}
m_{\text{eff},q}\simeq \pi \kappa\left(q+\frac{1}{4}\right)e^{-\pi \kappa R_0}, \;\;\;q\geq 1,
\end{equation}in \eqref{eq7_15_2} as in \cite{31}, we find that in the case of $DD$ boundary conditions, i.e., $\chi=1$, we obtain  the same result as derived in \cite{31} (formula (2.18)).
In the general case where there are $n$ extra dimensions compactified to a torus $T^n$ on the branes, with radius $R_1, \ldots, R_n$ respectively,
the authors in \cite{33} used the approximation
\begin{equation*}m_{\text{eff},q}\simeq \pi \kappa\left(q+\frac{1}{2}\right)e^{-\pi \kappa R_0},\;\;\;q\geq 1\end{equation*}which is only good if the internal manifold has dimension $n=1$. Using this approximation in \eqref{eq7_15_4} with $\chi=1$, it seems that there are still some discrepancies between the result of \cite{33} (see formula (32) in \cite{33}) and our result \eqref{eq7_15_4} in the terms corresponding to $q=0$. However, if one applies the Chowla-Selberg formula for Epstein zeta functions to the three terms on the first two lines of (32) in \cite{33}, one would recover the sum of the terms with $q=0$. Compared to the works of \cite{31,33}, our results \eqref{eq7_15_2} and \eqref{eq7_15_4} do not use any approximations to the effective masses. They are exact results. Moreover, our formulas \eqref{eq7_15_2} and \eqref{eq7_15_4} show manifestly that the Casimir force is attractive for Dirichlet-Dirichlet boundary conditions.

Since RSKKII model is a limit of the RSKKI model when the compactifying radius $R_0$ of the extra dimension with coordinate $y$ becomes infinite, one would expect that the same conclusions about the sign of the Casimir force still hold for the RSKKII model. From \eqref{eq7_13_2}, one can show that as $\kappa R_0\gg 1$,
 $
m_{\text{eff},q}, q\geq 1,
$ is approximately equal to $\pi \kappa e^{-\pi\kappa R_0}(q+\vep)$ for some $\vep$. This implies that $m_{\text{eff},q}, q\geq 1$ can be considered as the eigenvalues coming from  an extra dimension compactified to a (twisted) torus of radius $e^{\pi\kappa R_0}/(\pi \kappa)$. Therefore \eqref{eq7_9_17} shows that for $R_0$ large enough, the magnitude of the Casimir force is increased if we increase  the compactifying radius $R_0$. In the special cases considered in \cite{31,33}, this behavior was confirmed by the  figures in \cite{31,33}. In fact, the same argument as in \cite{29} shows that the Casimir force will be proportional to the radius $e^{\pi\kappa R_0}/(\pi\kappa)$ of the extra dimension. Therefore when all the parameters except $R_0$ are kept fixed and $R_0\rightarrow \infty$, the Casimir force \eqref{eq7_9_17} is proportional to $e^{\pi\kappa R_0}$, and if $R_0$ goes to infinity in such a way that $\kappa R_0$ is kept fixed, then the Casimir force \eqref{eq7_9_17} is proportional to $R_0$. As a result, we cannot take the limit $R_0\rightarrow \infty$ on the Casimir force \eqref{eq7_9_17} directly.   In \cite{31,33}, finite results have been claimed for the Casimir force density for the RSII model. In these articles, the authors retained the $q=0$ terms in \eqref{eq7_15_2} or \eqref{eq7_15_4}, and change the summation over $q\geq 1$ to an integral over $m$. The latter is tantamount to dividing by the factor $e^{\pi\kappa R_0}$ and taking the limit $R_0\rightarrow \infty$ on the summation of $q\geq 1$ terms. This combination of retaining the $q=0$ term and changing the summation to integral  is not equivalent to the direct $R_0\rightarrow \infty$ limit. As we have discussed, one should expect that in the limit $R_0\rightarrow \infty$, the Casimir force density acting on infinite parallel plates increases beyond all bounds. Nevertheless, it is easy to see that the procedure used in \cite{31,33} gives Casimir force  that is attractive for homogeneous boundary conditions and repulsive for mixed boundary conditions.

As is discussed earlier, the presence of the extra dimensions enhances the Casimir force.  It will be interesting to study whether the extra dimensions give a significant increase to the Casimir force. It is sufficient to consider the case with $DD$ boundary conditions. We discuss first the case of the RSI model where there is no internal space. At any temperature, numerical computations of \eqref{eq7_15_1} and \eqref{eq7_15_4} show that for plate separation $a$ in the range $100$nm $\sim$ $ 1000$nm, the correction to the Casimir force in $(3+1)$-D spacetime is less than $0.1\%$ for $\kappa e^{-\pi\kappa R_0}>$3\,eV. For $a=100$nm, the correction to the Casimir force becomes $\sim 10\%$ when $\kappa e^{-\pi\kappa R_0}\sim $ 1.40\,eV; and for $a=1000$nm, the correction to the Casimir force becomes $\sim 10\%$ when $\kappa e^{-\pi\kappa R_0}\sim $ 0.13\,eV. As is well known, the resolution of the hierarchy problem requires $\kappa R_0\sim 12$. In this case, the correction to the Casimir force is significant if $\kappa \sim 10^7$\, GeV. If $\kappa$ is of Planck scale $\sim 10^{19}$\,GeV, then $\kappa e^{-\pi\kappa R_0}\sim \,$400\,GeV $\gg$ 3eV, and the correction to the Casimir force would be too small to be observed in the current Casimir experiments. This situation can be  changed if there exists an internal space that has size comparable to the plate separation. Assume that the internal space is $S^1$ with radius $R$. In the absence of the extra dimension $S^1/\mathbb{Z}_2$ or when $\kappa e^{-\pi\kappa R_0}\gg $ 3eV, the correction to the Casimir force is less than $0.1\%$ if $R/a<0.1$. If $R/a\sim 0.3$, then the correction will grow to $\sim 10\%$. When $R/a\sim 1$, the correction is $\sim 200\%$. However, it is believed that the internal manifold should be compactified to a size much smaller than $1$nm, therefore the existence of internal space would not be able to be detected by the present Casimir experiments which measures Casimir force for separations larger than $1$nm.
 \begin{figure}[h]\centering \epsfxsize=.49 \linewidth
\epsffile{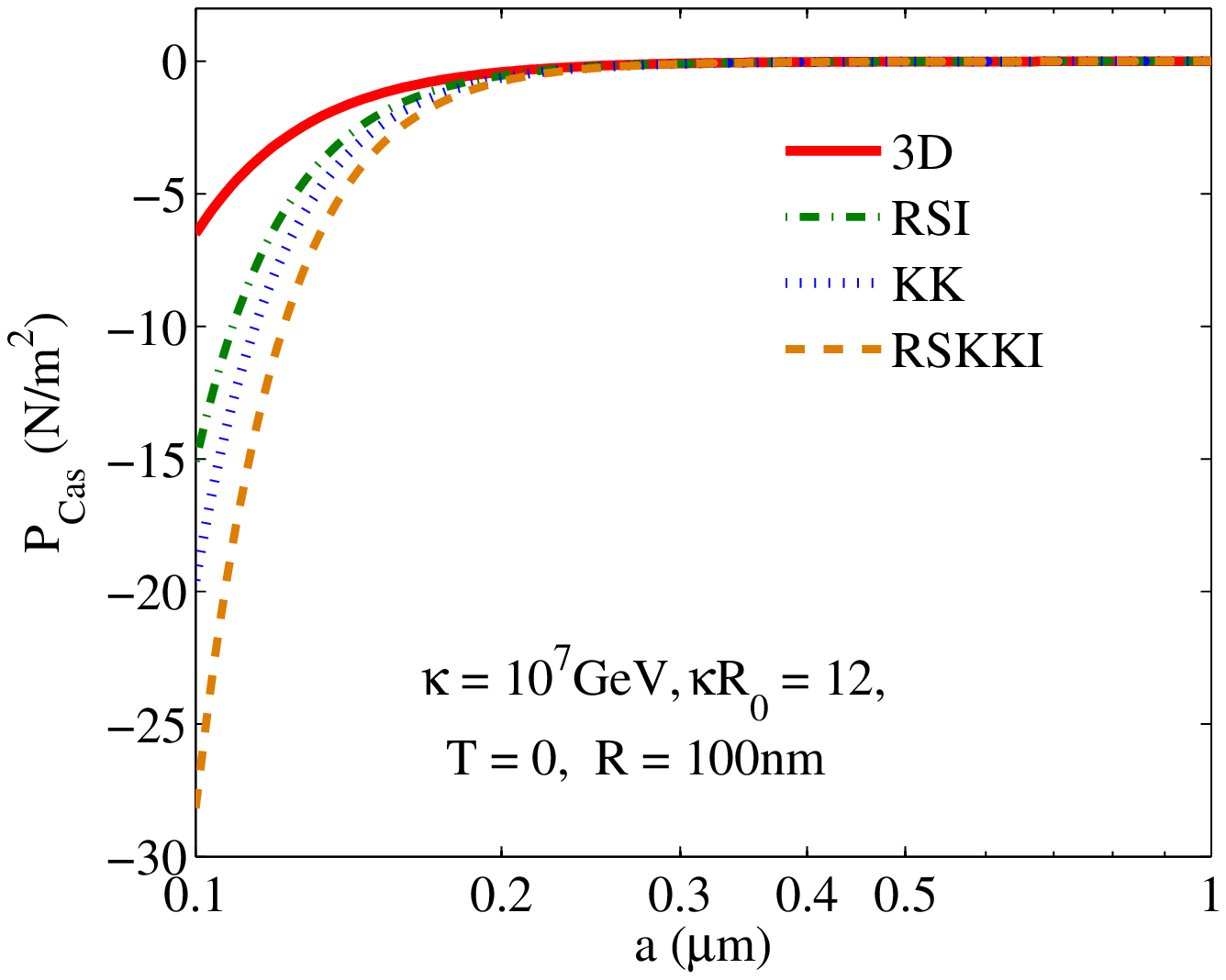}\centering \epsfxsize=.49 \linewidth
\epsffile{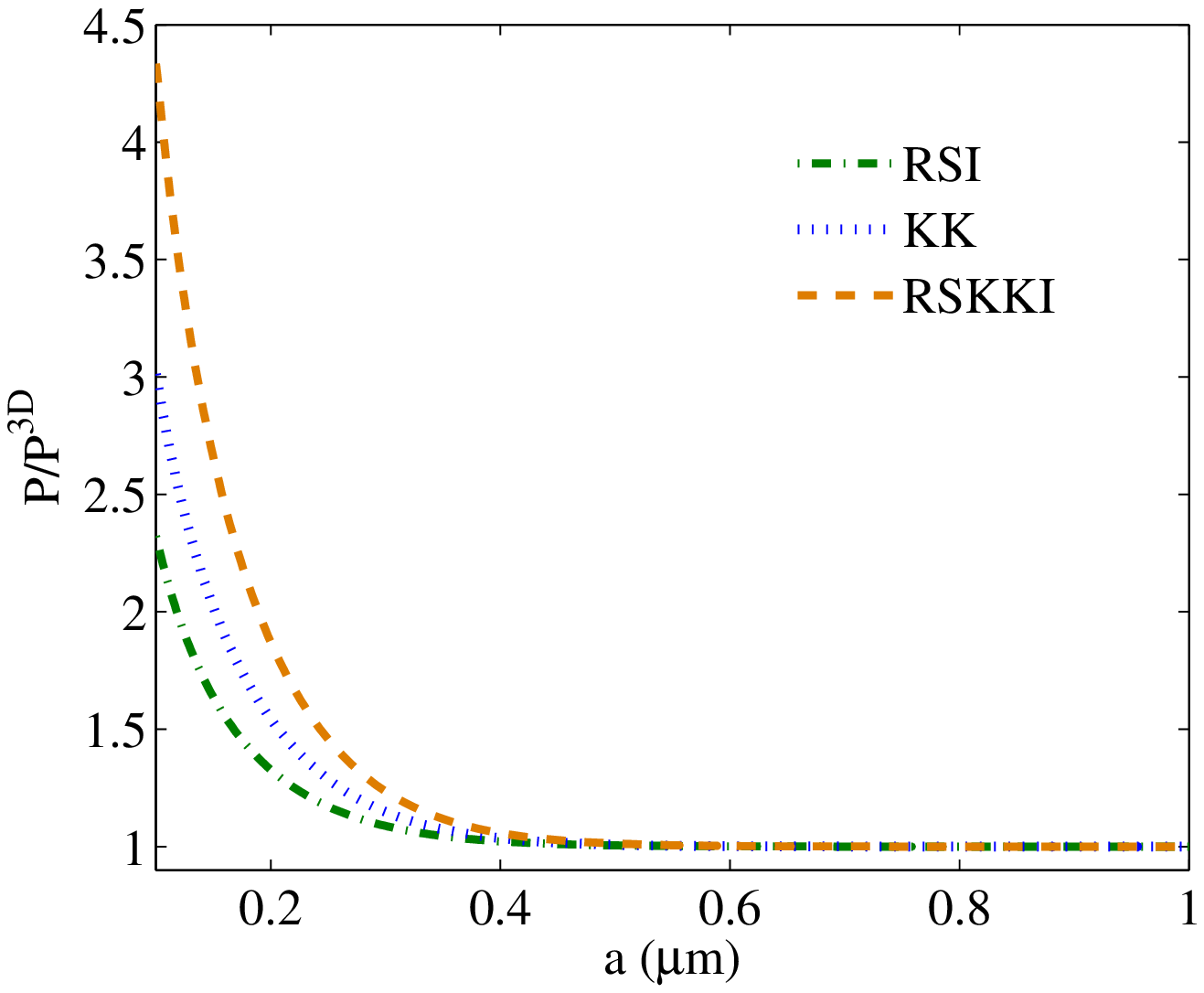}\caption{\label{f2} These graphs compare the zero temperature Casimir pressures  for $DD$ boundary conditions, in (3+1)-D spacetime, in RSI model with $\kappa =10^7$GeV and $\kappa R_0=12$, in KK spacetime with internal manifold a circle of radius $R=100$nm, and in RSKKI model. The graph on the right shows the ratio of the pressures to the pressure in $(3+1)$-D.}\end{figure}
In FIG. \ref{f2} and FIG. \ref{f3}, we show the Casimir pressures \eqref{eq7_15_1} and \eqref{eq7_15_4} for $DD$ boundary conditions in $(3+1)$-dimensional Minskowski spacetime, in RSI spacetime with $\kappa = 10^7$\,GeV and $\kappa R_0=12$, in KK spacetime where the internal manifold is a circle with radius $R=100$nm, and in RSKKI spacetime, at zero temperature and at temperature $T=1$\,MeV ($\sim 10^{10}$ K) respectively. The parameters are chosen so that there are significant differences between the Casimir pressures in various spacetimes. Compare FIG. \ref{f2} and FIG. \ref{f3}, we see that high temperature has profound effect on the strength of the Casimir force, as dictated by the linear dependence of the Casimir force in temperature in the high temperature regime.

 \begin{figure}[h]\centering \epsfxsize=.49 \linewidth
\epsffile{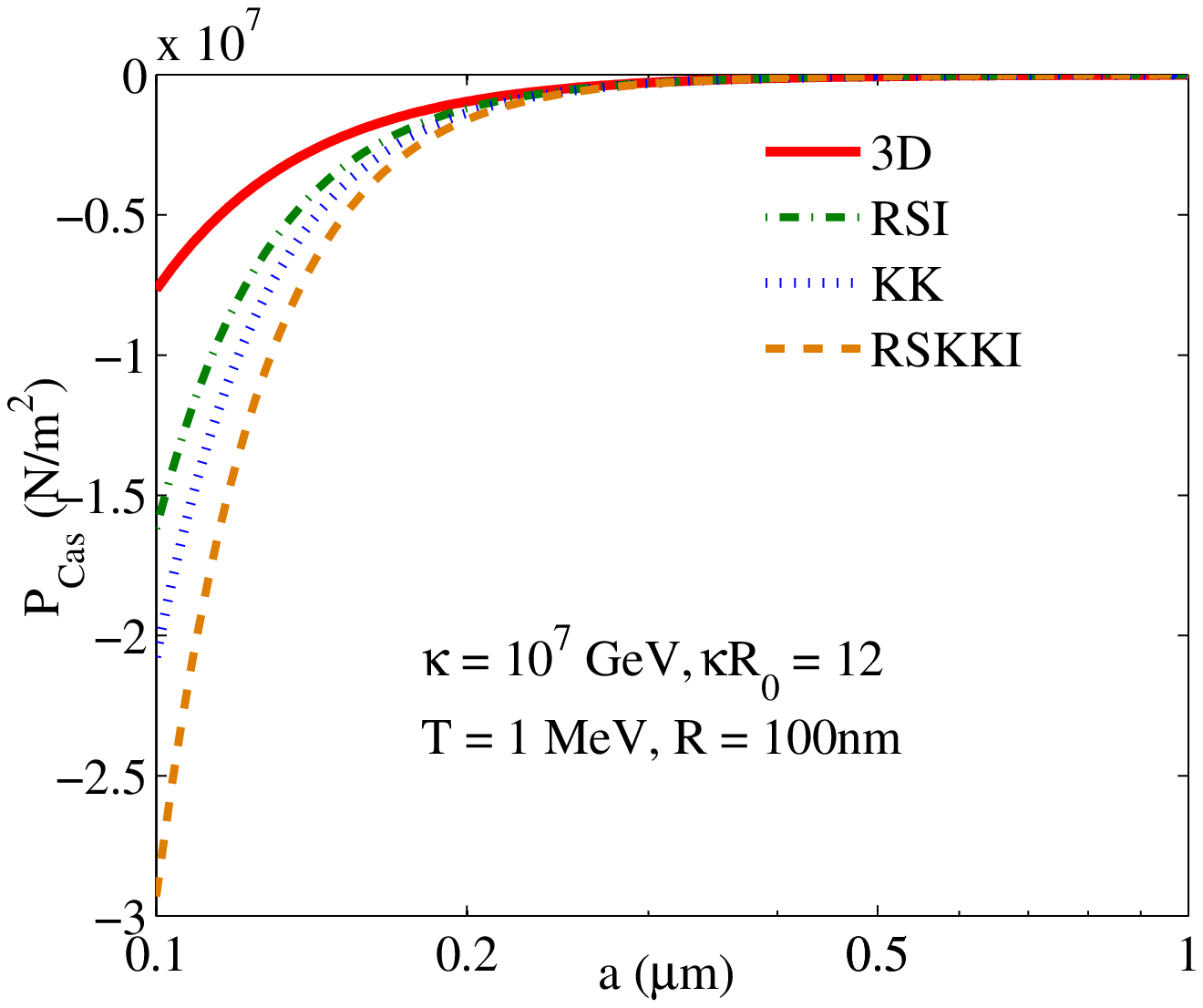}\centering \epsfxsize=.49 \linewidth
\epsffile{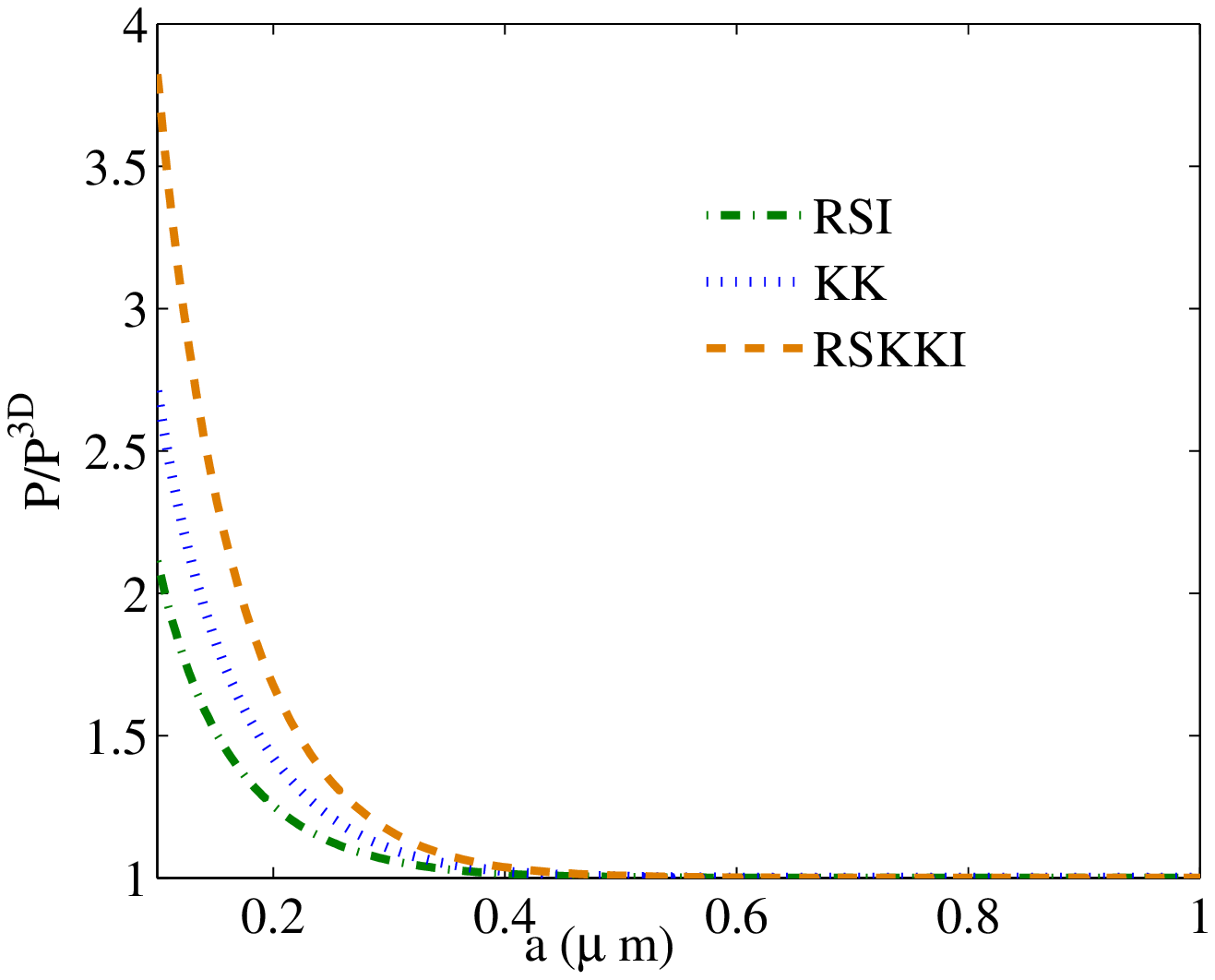}\caption{\label{f3} Same as FIG. \ref{f2} but with $T=1$MeV.}\end{figure}
\section{  Conclusions}We have given a brief discussion about the Casimir force acting on a pair of parallel plates in higher dimensional warped spacetime model which are generalizations of $(4+1)$-D Randall-Sundrum   spacetime model.  The contributions of this article are the followings. We have derived exact formulas for the Casimir force acting on a pair of parallel plates due to a massless scalar field without using any approximations. From the exact formulas, we showed that the Casimir force is always attractive for Dirichlet-Dirichlet or Neumann-Neumann boundary conditions, and repulsive for Dirichlet-Neumann conditions. Although the discussions in this article are restricted to the extension of Randall-Sundrum model, which we call Randall-Sundrum-Kaluza-Klein model, and   specific boundary conditions have been  imposed on the branes,    the discussions in this article can be applied in a more general context. Notice that the change in the boundary conditions on the branes only alters the spectrum of the effective masses, and the change in the geometry of the extra dimensions only alter the spectrum of the internal manifold.  Therefore the results of this article can be applied to more general spacetime model with extra dimensions. One can deduce as in this article   that in the absence of  tachyonic modes,  the sign of the Casimir force due to a scalar field acting on a pair of parallel plates in the macroscopic $(3+1)$-dimensional Minskowski spacetime only depends on the boundary conditions imposed on the plates, and is not changed by the presence of extra dimensions. The extra dimensions only enhance the magnitude of the Casimir force.

It should  be mentioned that there is a recent work \cite{44} that proposed a different perspective on the Casimir force acting on parallel plates in spacetime with extra dimensions. The approach considered in this article  permits the field to propagate in the bulk, and we are in fact considering the Casimir force acting on a codimension one hyperplane. In \cite{44}, it was shown that if the plates are localized on the visible brane, the correction to the Casimir force should be much smaller. Finally, we would also like to mention that this article is a revised version of the preprint \cite{60}. A recent interesting work \cite{61} generalized our work \cite{60} and discussed the Casimir effect on parallel plates in the usual $(4+1)$-D RS model, but with general Robin boundary conditions on the plates. In case of $DD, NN$ or $DN$ boundary conditions, the authors of \cite{61} confirmed that they obtained the same result as our eq. \eqref{eq7_15_1}.

\begin{acknowledgments}The author would like to thank A. Flachi for explaining his work to us. We would also like to thank the anonymous referee for the helpful comments.
This project is   funded by Ministry of Science, Technology and Innovation, Malaysia under e-Science fund 06-02-01-SF0080.
\end{acknowledgments}

\end{document}